\documentclass[twocolumn,epsfig,aps,prb]{revtex4}

\begin{document}

\title{Enhanced transmission of light through a gold film due to excitation of standing surface plasmon Bloch waves.}

\author{Igor I. Smolyaninov, Yu-Ju Hung}
\affiliation{Department of Electrical and Computer Engineering, University of Maryland, College Park, MD 20742, USA}

\date{\today}

\begin{abstract}
We have observed enhanced transmission of light through a gold film due to excitation of standing surface plasmon Bloch waves in a surface Fabry-Perot resonator. Our experimental results strongly contradict the recently suggested model of light transmission via excitation of a composite diffractive evanescent wave.
\end{abstract}


\maketitle

Transmission of light through nanostructured metal films remains a topic of considerable interest and unresolved controversy over the last few years. Immediately after the first observations in the late nineties \cite{1,2} the dominant mechanism of light transmission was believed to involve excitation of surface plasmon polaritons (SPP) \cite{3,4} on the interfaces of the metal film.  Since the SPPs of the nanostructured metal interfaces are strongly coupled to each other and to the free space photons, they seem to be responsible for the effect of extraordinary optical transmission \cite{5}. Very recently a reassessment of the earlier transmission measurements of ref.\cite{1} and a new theoretical model of the enhanced optical transmission has been offered \cite{6,7}, which relies on a new theoretical construction called a composite diffractive evanescent wave (CDEW). The main claim of these recent papers is that SPPs play a minor auxiliary role in the optical transmission of nanostructured metal films, while excitation of CDEWs is primarily responsible for the effect of extraordinary light transmission. The main difference between SPPs and CDEWs is that the SPP is a propagating long-range surface wave of charge density and the associated electromagnetic field \cite{3,4}, while according to ref.\cite{6} CDEW is a combination of diffracted evanescent modes originating at an abrupt surface discontinuity such as a hole, a bump, etc. The wavelength of a CDEW corresponds to the wavelength of the excitation light in free space \cite{6}. The CDEW is not a propagating surface wave in a sense that its field intensity $E^2$ is proportional to $1/x^2$, where $x$ is the distance from the source \cite{6}. Since many practical devices are being built or are proposed to be built using the effect of extraordinary light transmission through metal films, understanding of the main mechanism behind this effect is very important. At the moment, the discussion about the comparative role of SPPs and CDEWs in the effect in question is conducted primarily based on theoretical arguments. In this letter we report a set of simple experiments, which strongly favor the dominant role of SPPs in the effect of extraordinary light transmission through nanostructured metal films. We show that transmission of light through a gold film is strongly enhanced due to excitation of standing SPP Bloch waves in a surface Fabry-Perot resonator, which is approximately 16 micrometers long. Our experimental results strongly contradict the model of light transmission via excitation of a CDEW. Another important implication of our results is that ability to make two-dimensional Fabry-Perot resonators for surface plasmon polaritons brings us closer to experimental realization of a SPASER \cite{8}, a quantum coherent source of SPPs. A plasmonic medium with gain has been already demonstrated recently \cite{9}. Thus, in order to realize the SPASER we need to bring these two advances together in one device.
    
In our experiments 50 nm thick gold films were sputtered onto a glass substrate using a Magnetron Sputtering Machine. An overlay of polymethyl methacrylate (PMMA) film was then spin-coated and patterned using E-beam lithography. The dielectric PMMA film was about 100-200 nm thick. An example of a patterned bi-grating is shown in Fig. 1(a). It consists of individual PMMA dots on gold film surface. The grating period is 500 nm in both directions. The gold films under the PMMA layer were still intact after the gratings were developed using MIBK/IPA developer. With 502 nm laser light illumination the area of the gold film above which the PMMA grating has been formed exhibited the effect of extraordinary optical transmission similar to the one described in ref.\cite{1}. This is illustrated in Figs.1(b,c), in which the area of the gold film under the PMMA grating appears similar in brightness to the areas of the gold film punctured with a periodic array of 200 nm diameter nanoholes when illuminated with the same laser power.           

The idea of our experiment was to fabricate a Fabry-Perot type surface resonator for the SPP Bloch waves and to check if there would be a noticeable increase in the effect of extraordinary light transmission associated with the resonance. Note that since CDEWs are non-propagating waves, this effect would be impossible to observe with CDEWs in a surface Fabry-Perot resonator, which is a few tens of micrometers long. 

A surface plasmon polariton Bloch wave can be excited on the surface of the rectangular bi-grating (Fig.1(a)) at a number of resonant angles. These angles $\alpha _n$ are determined by the quasi-momentum conservation law \cite{4}: 

\begin{equation}
\label{eq1}
k\sin \alpha _n=k_p+2\pi n/a ,
\end{equation}

where $k$ is the wavevector of photons in free space, $k_p$ is approximately equal to the SPP wave vector on the flat metal surface, $a$ is the period of the structure in the direction of propagation, and $n$ is an integer. Such SPP Bloch waves have been observed in ref.\cite{10}. The geometry of our sample is shown in Figs.2(a,c). A corner of the rectangular array of the PMMA dots shown in Fig.1(a) occupies the bottom left corner of Fig.2(a). The outside area is the non-strucutered PMMA film on top of the gold film. The T-shaped feature inside the array of PMMA dots was exposed to the E-beam and etched away, so that the gold film is naked in this area. The AFM image of the edge of the PMMA dots array is shown in Fig.2(c). The microscopic photo image of this structure illuminated by 532 nm laser light is shown in Fig.2(b). It is quite apparent from this image that the area of the PMMA dot array between the end of the long leg of the T and the nearest edge of the PMMA dot array exhibits much higher transmission than the rest of the PMMA dot array. On the other hand, the two areas of the array between the edge of the structure and the two arms of the T exhibit about the same transmission as the rest of the PMMA dot array. The observed effect is very sensitive to the illumination angle. Compared to the image in Fig.2(b), the illumination angle of the 532 nm laser light in Fig.2(d) was increased by only 3 degrees. 

In order to prove that the area between the leg of the T and the nearest edge of the PMMA array behaves as a Fabry-Perot resonator for surface plasmon polaritons we have analyzed the two cross sections of the image in Fig.2(b). These cross sections are shown in Fig.3. The directions of the cross sections are shown in Fig.3(a). The cross section in Fig.3(b) is performed along the optical axis of the Fabry-Perot resonator. The maxima of the optical field indicated by the arrows are separated by $\Lambda _{Bloch}/2$ distances, where the wavelength of the SPP Bloch wave 

\begin{equation}
\label{eq2}
\Lambda _{Bloch}=-\frac{a\lambda _p}{(\lambda _p-a)}\approx 8.3 \mu m
\end{equation}

corresponds to $n=-1$ in equation (1). The second cross section in Fig.3(c) is performed along a diagonal direction. It indicates that the intensity of light transmitted inside the plasmonic Fabry-Perot resonator is enhanced by approximately a factor of 4 compared to the transmission of the rest of the PMMA dot array. It appears that this result strongly contradicts the CDEW model described in refs.\cite{6,7}. The intensity of a CDEW wave would drop by a factor of $(2\mu m/32\mu m)^2\approx 0.004$ upon the double pass through the Fabry-Perot interferometer. This estimate is based on the $\approx 2\mu m$ CDEW propagation length cited in ref.\cite{6}, and the length of the plasmonic Fabry-Perot resonator shown in Figs.2,3. Thus, no resonance enhancement of the effect of the extraordinary light transmission is possible due to excitation of CDEW in the resonator. On the other hand, the resonator length is smaller than the typical values of the SPP propagation length measured in the experiments and calculated theoretically \cite{2,4}, while the length of the resonator corresponds to approximately two SPP Bloch wavelengths estimated using equation (2). We believe that the plasmon-assisted mechanism of light transmission is proven beyond reasonable doubt.   

In conclusion, we have observed enhanced transmission of light through a gold film due to excitation of standing surface plasmon Bloch waves in a surface Fabry-Perot resonator. Our experimental results strongly contradict the recently suggested model of light transmission via excitation of a composite diffractive evanescent wave. Another important implication of our results is that ability to make two-dimensional Fabry-Perot resonators for surface plasmon polaritons brings us closer to experimental realization of a SPASER \cite{8}.

This work has been supported in part by the NSF grants ECS-0304046, CCF-0508213, and ECS-0508275.

\begin{figure}
\begin{center}
\end{center}
\caption{ (a) An AFM image of a bi-grating fabricated on the surface of 50 nm thick gold film, which consists of individual PMMA dots. The grating period is 500 nm in both directions. (b) With 502 nm laser light illumination the area of the gold film above which the PMMA grating has been formed exhibits the effect of extraordinary optical transmission similar to the one described in ref.\cite{1}. This is illustrated in (c), in which the area of the gold film punctured with a periodic array of 200 nm diameter nanoholes appears to have similar brightness when illuminated with the same laser power.}
\end{figure}

\begin{figure}
\begin{center}
\end{center}
\caption{ Photos of the Fabry-Perot resonator for the SPP Bloch waves illuminated from the top with white light (a), and illuminated from the bottom with 532 nm light (b). (c) shows an AFM image of the boundary between the non-structured area of the PMMA film on top of the gold film and the area of the gold film covered with a periodic array of PMMA bumps. (d) The effect of resonant transmission disappears when illumination angle is increased by 3 degrees: compare (b) and (d).}
\end{figure}

\begin{figure}
\begin{center}
\end{center}
\caption{The lines in (a) indicate the directions of the cross sections of the optical intensity shown in (b) and (c). The cross sections are made along the optical axis of the resonator (b), and in the diagonal direction (c). The distance between the maxima of the optical intensity in (b) corresponds to $\Lambda _{Bloch}/2$. The cross-section in (c) indicates that the transmission of the PMMA dot array inside the Fabry-Perot resonator is approximately four times higher than in the rest of the PMMA dot array.}
\end{figure}

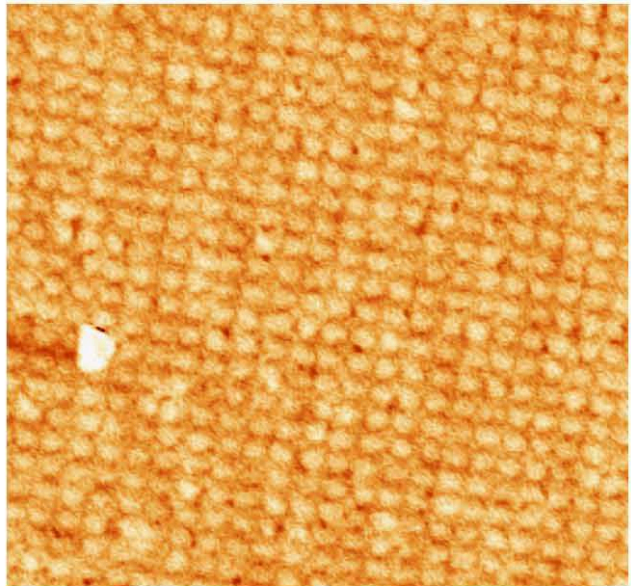
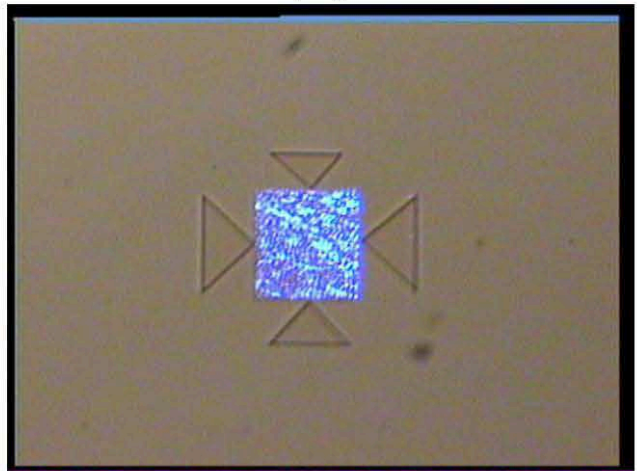
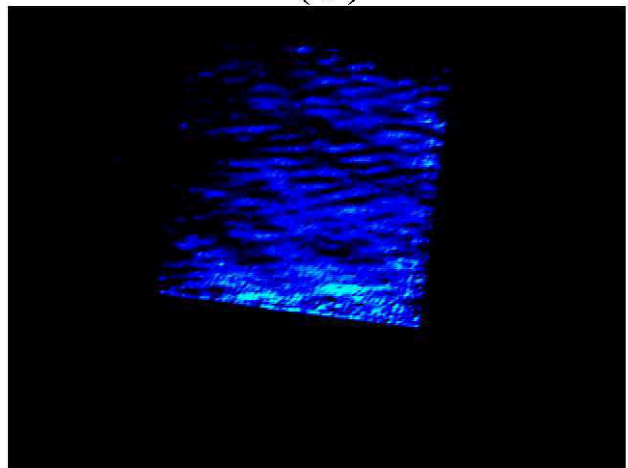
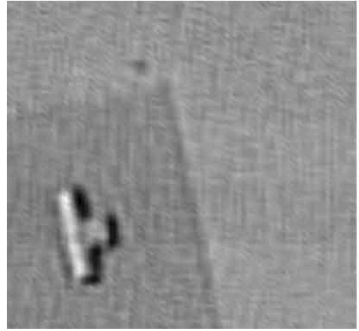
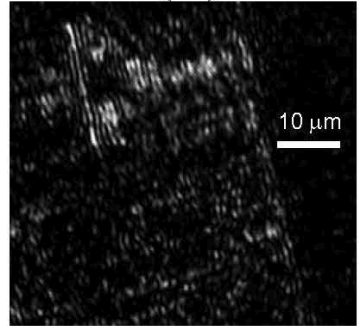
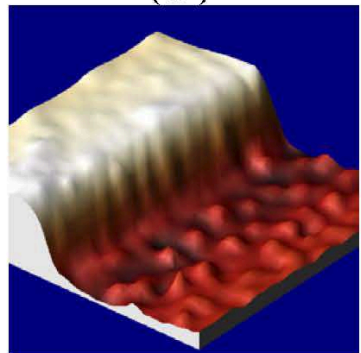
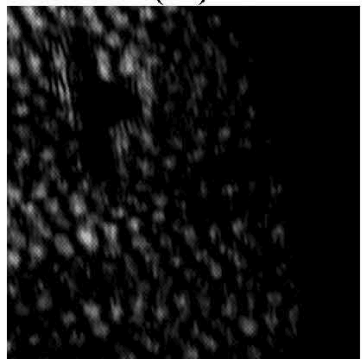
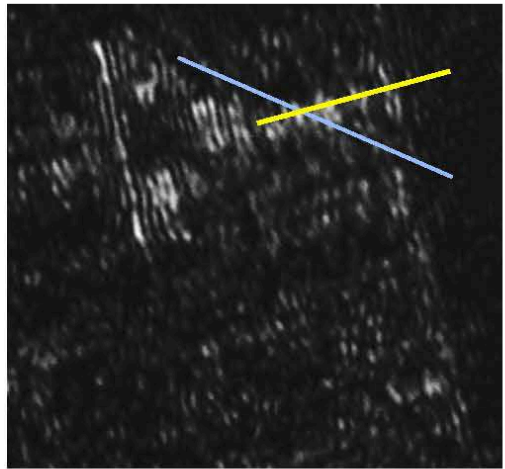
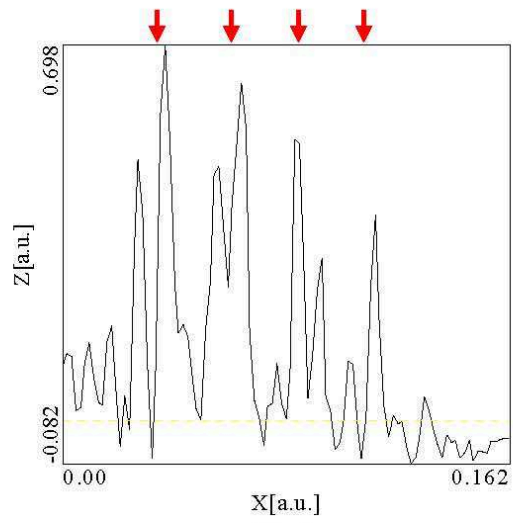
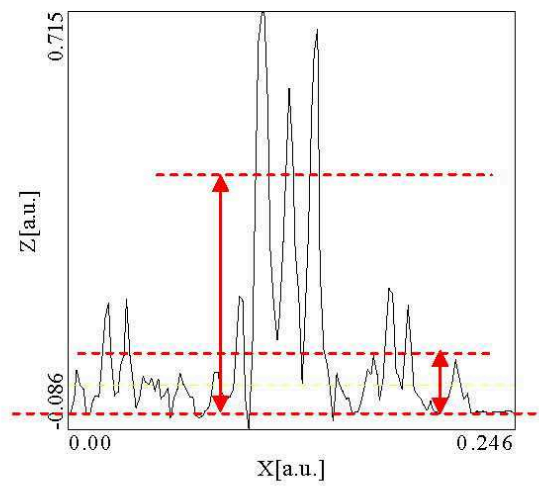
\end{document}